\documentclass[aip,
 jmp,
 amsmath,amssymb,
 reprint,%
]{revtex4-1}

\usepackage{graphicx}
\usepackage{dcolumn}
\usepackage{bm}
\usepackage{enumerate}

\begin{document}


\title{A study of adatom ripening on an Al (111) surface with machine learning  force fields}

\author{V. Botu}
 \email{botuvenkatesh@gmail.com}
\author{J. Chapman}
\author{R. Ramprasad}%
\affiliation{ 
University of Connecticut, Storrs, CT 06269
}%

\begin{abstract}
Surface phenomena are increasingly becoming important in exploring nanoscale materials growth and characterization. Consequently, the need for atomistic based simulations is increasing. Nevertheless, relying entirely on quantum mechanical methods limits the length and time scales one can consider, resulting in an ever increasing dependence on alternative machine learning based force fields. Recently, we proposed a machine learning approach, known as AGNI, that allows fast and accurate atomic force predictions given the atom's neighborhood environment. Here, we make use of such force fields to study and characterize the nanoscale diffusion and growth processes occurring on an Al (111) surface. In particular we focus on the adatom ripening phenomena, confirming past experimental findings, wherein a low and high temperature growth regime were observed, using entirely molecular dynamics simulations.
\end{abstract}

\maketitle

As the fabrication of materials continually progresses towards the atomic-scale, an interest in layer by layer growth methods (such as molecular-beam epitaxy or atomic layer deposition), in micro-electronics, catalysis, or biomedical applications, has risen tremendously.\cite{Johnson_1,Steven_1,Booyong_1,Herman_1,Pamplin_1,Zhang_1} The high degree of control offered allows for a sub-nanometer scale precision in the morphological structure of the materials grown. Consequently, the need to better understand and characterize such growth processes, at the atomic level, has emerged.

Towards this cause, the advent of first-principles (also known as \textit{ab initio}) based \emph{in silico} models has been instrumental. Methods such as density functional theory, along with harmonic transition state theory, are now commonly used to (i) map out the energetics for the constitutive elementary reaction pathways, and (ii) then, rely on coarser stochastic approaches (e.g. kinetic Monte Carlo), to spatially and temporally evolve the state of a system, thus, helping unravel the complex atomistic growth phenomena at significantly larger length and time scales.\cite{Elliott_1, Voter_1}  Nevertheless, building a complete catalog of reaction pathways \textit{a priori}  is often challenging and non-trivial for low symmetry systems. An alternative, and more natural, formalism is to use molecular dynamics simulations, whereby the temporal state of an atomistic system is evolved by solving Newton's equations of motion. The key ingredient required for such methods is a description of the forces between the interacting atoms. Two methods - quantum mechanics or semi-empirical potentials, allow access to these forces. Unfortunately, the formidable computational cost of quantum mechanical methods restricts the time and length scales one can consider, while semi-empirical approaches provide a cheaper alternative but often lack the versatility and accuracy of quantum mechanical interactions. If pathways to accelerate \textit{ab initio} methods whilst retaining accuracy existed, they would be highly desired.
%

Off late, the prominence of machine learning methods, when used in tandem with quantum mechanical generated data, has demonstrated to be a reasonable trade-off between the cost, accuracy, and versatility issues facing current methods.\cite{Bartok_2,Behler_1,Behler_2,Botu_1,Botu_2,Li_1} Recently, we introduced one such framework (known as AGNI force fields), wherein, by encoding an atom's neighborhood environment numerically (its fingerprint), a mapping to the vectorial force it experiences can be established.\cite{Botu_1,Botu_2} Such interpolative force fields were shown to be able to retain quantum mechanical accurate force predictions, with errors $<$ 0.05 eV/\AA. Further, the high fidelity framework now comes at a fraction of the cost compared to quantum mechanical methods. 

This letter is intended to demonstrate the use of AGNI force fields in exploring nanoscale growth phenomena. In particular, we study the ripening of adatoms, whereby individual atoms cluster together to form larger island features, that occurs during the layer by layer growth process for an Al (111) surface, using true molecular dynamics (MD) simulations. This goes well beyond our previous work where we studied the dynamical behavior of a single adatom on the same surface.\cite{Botu_2} The (111) surface was chosen in particular as the barriers for the elementary processes  allow for the ripening phenomena to be explored in time scales achievable with conventional MD simulations. The remainder of the letter is structured as follows. First, a brief review on the construction of AGNI force fields is provided. The force field is then validated by computing energetic barriers for elementary reaction pathways encountered on the surfaces. We then perform long time scale MD simulations exploring the ripening process as a function of two important parameters, time and temperature.

\begin{figure}
	\centering
	\includegraphics[scale=0.3]{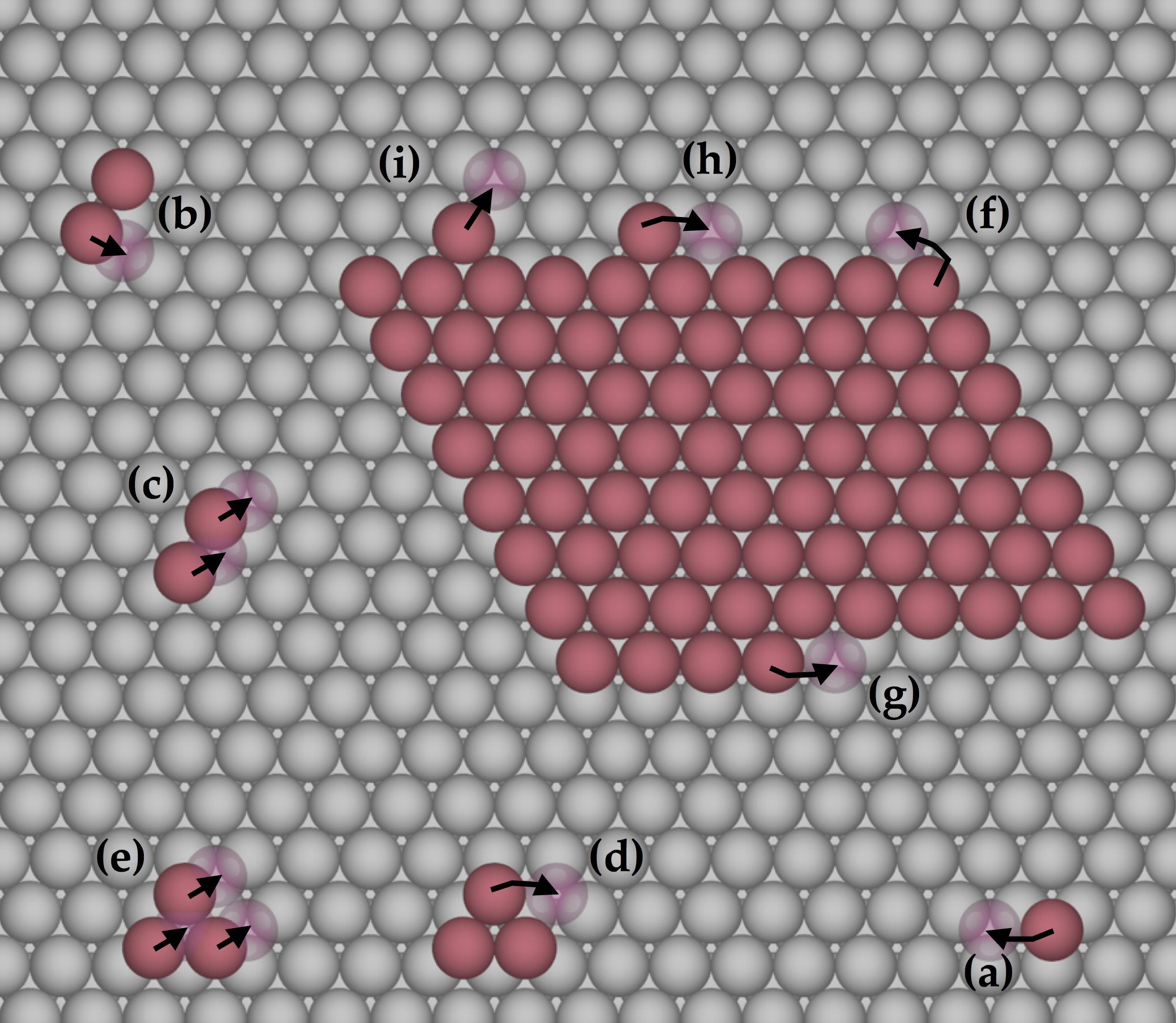}
	\caption{Elementary reaction pathways of monomer, dimer, trimer, and other island features on an Al (111) surface that lead to the ripening phenomena. Grey and red colored atoms correspond to the surface atoms and adatoms, respectively.} 
	\label{pathways}
\end{figure}

\begin{table}
	\caption{Activation energy barriers for reaction pathways plotted in Figure \ref{pathways}, computed with AGNI (E$_a^{AGNI}$) and DFT. E$_a^{AGNI}$ were obtained by integrating the forces. E$_a^{DFT}$ were computed using the climbing-image nudged elastic band method, or sourced from literature as indicated by *.\cite{Bogicevic_1,Bogicevic_2}} \label{Table:Barriers} 
	\begin{tabular}{ccccccc}
		\hline
		\hline
				\multicolumn{1}{c}{\textbf{Pathway}} & \multicolumn{3}{c}{E$_a^{AGNI}$} & \multicolumn{3}{c}{E$_a^{DFT}$} \\ 
		\hline
		\multicolumn{1}{l}{} & \multicolumn{3}{c}{} & \multicolumn{3}{c}{} \\
		\multicolumn{1}{l}{(a) Monomer hopping} & \multicolumn{3}{c}{0.05} & \multicolumn{3}{c}{0.04} \\
		\multicolumn{1}{l}{(b) Dimer rotation} & \multicolumn{3}{c}{0.12} & \multicolumn{3}{c}{0.11} \\
		\multicolumn{1}{l}{(c) Dimer translation} & \multicolumn{3}{c}{0.13} & \multicolumn{3}{c}{0.07} \\
		\multicolumn{1}{l}{(d) Trimer translation} & \multicolumn{3}{c}{0.19} & \multicolumn{3}{c}{0.21} \\
		\multicolumn{1}{l}{(e) Trimer rotation} & \multicolumn{3}{c}{0.19} & \multicolumn{3}{c}{0.24} \\
		\multicolumn{1}{l}{(f) Corner evaporation} & \multicolumn{3}{c}{0.71} & \multicolumn{3}{c}{0.60*} \\
		\multicolumn{1}{l}{(g) Kink evaporation} & \multicolumn{3}{c}{0.67} & \multicolumn{3}{c}{0.65*} \\
		\multicolumn{1}{l}{(h) Edge diffusion} & \multicolumn{3}{c}{0.48} & \multicolumn{3}{c}{0.45*} \\
		\multicolumn{1}{l}{(i) Edge evaporation} & \multicolumn{3}{c}{0.91} & \multicolumn{3}{c}{0.80*} \\		
		
		\hline
	\end{tabular}
\end{table}

\begin{figure*}
	\centering
	\includegraphics[scale=0.4]{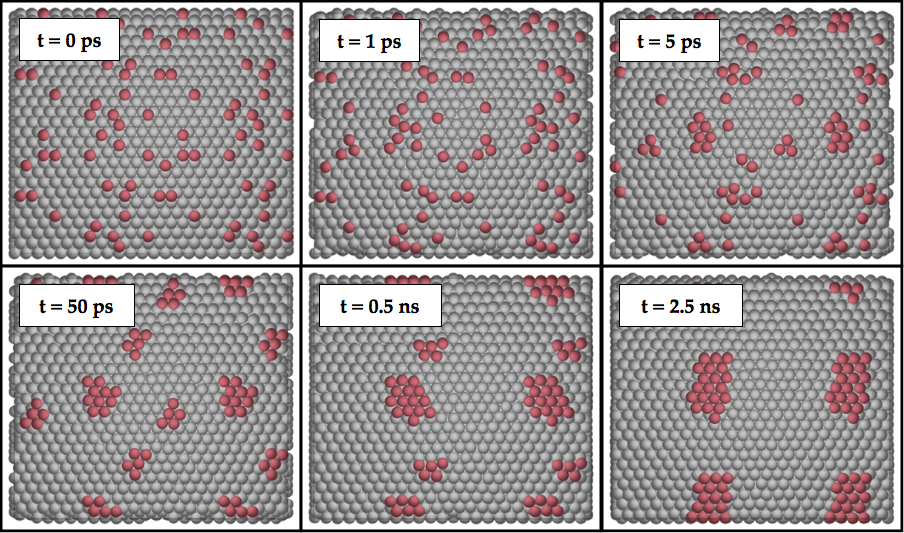}
	\caption{Snapshots of the time-evolution of adatoms on Al (111) surface using constant temperature (300 K) molecular dynamics simulation. Adatoms were randomly distributed on the surface as shown at t = 0 ps, $\theta$ = 0.14. The images shown are a 2 $\times$ 2 repeat of the unit cell. Grey and red colored atoms correspond to the surface atoms and adatoms, respectively.} 
	\label{ripening_time}
\end{figure*}

The construction of an accurate AGNI force field consists of 4 primary steps: (i) generating a diverse dataset of quantum mechanical reference atomic environments and forces, (ii) choosing a numerical representation for the atomic environments, (iii) down selecting a representative set of training atomic environments, and (iv) identifying a learning method to map the environment to the force. For a detailed account of this workflow we refer the reader to our recent contributions.\cite{Botu_1,Botu_2,Botu_3} Here, we only provide a brief overview of the method. We begin by compiling periodical and non-periodical atomic configurations of Al, e.g., bulk, surfaces, clusters, defects, etc. For each configuration, \textit{ab initio} based MD simulations, based on density functional theory (DFT), were performed at multiple temperatures to compile a diverse set of equilibrium and non-equilibrium atomic environments, along with their corresponding vectorial forces. All the DFT calculations were run using the VASP code at a PBE level of theory.\cite{Kresse1,Kresse2,Blochl1,Car_1} Each atom and its environment is then encoded in a numerical manner using a directional dependent representation. The exact functional form is given below as

\begin{equation}\label{eq:atom}
V_i^u(\eta) = \sum\limits_{j \neq i} \frac{r_{ij}^u}{r_{ij}} \cdot e^{-{\left(\frac{r_{ij}}{\eta}\right)}^2} \cdot  f_d{(r_{ij})}.
\end{equation}

\noindent Here, $r_{ij}$ is the distance between atoms $i$ and $j$ ($||\mathbf{r}_j - \mathbf{r}_i||$), while $r_{ij}^u$ is a scalar projection of this distance along a direction $\hat{u}$. $\eta$ is the Gaussian function width. Multiple such $\eta$ values are needed to accurately describe the environment around an atom. Here, 8 $\eta$ values were sampled on a logarithmic grid between [0.8\AA, 16\AA]. $f_d(r_{ij}) = 0.5\left[\cos\left(\frac{\pi r_{ij}}{R_c}\right)+1\right]$, is a damping function for atoms within the cutoff distance ($R_c$), and is zero elsewhere. An $R_c$ of 8 {\AA} is used in the present work. From all the atomic environments compiled, a subset of these are chosen with the help of dimensionality reduction methods and sampling techniques, to ensure that the information content within these environments is diverse. Here 3000 such environments are chosen. Lastly, to map the atomic fingerprints to the forces we rely on the non-linear kernel ridge regression framework, along with standard machine learning practices to avoid over-fitting of the data. The constructed AGNI force field has a mean absolute prediction error (MAE) $<$ 0.05 eV/\AA, on the order of expected chemical and numerical accuracy of the reference quantum mechanical calculations.   

Moving on, a first step to the realization of the ripening process is to ensure that the elementary reactions occurring on the surface, such as translation, rotation and diffusion of adatoms, monomer, dimer, trimer and beyond, as well as processes corresponding to re-arrangement of islands, such as corner breaking, kink breaking, terrace diffusion, edge evaporation etc., as illustrated in Figure \ref{pathways}, are correctly described. Here, we compute the reaction energy barriers for all pathways, (a)-(i), shown in Figure \ref{pathways} using the AGNI force fields (E$_a^{AGNI}$). The barriers are reported in Table \ref{Table:Barriers}. Given that AGNI force fields provide access to the forces only we compute the energies via thermodynamic integration of the forces.\cite{Botu_3} For comparison we report the corresponding DFT computed reaction barriers (E$_a^{DFT}$), using the climbing-image nudged elastic band method and those reported in the literature (highlighted by * in Table \ref{Table:Barriers}).\cite{Bogicevic_1,Bogicevic_2} The errors are within 5\% of the DFT computed values. In principle, the accuracy can be improved upon by directly including these atomic environments during force field training, though such an undertaking is not considered here. 

 
Having demonstrated an accurate description of the elementary processes we now study the ripening of adatoms on the Al (111) surface. To do so we construct a 35 $\times$ 30 \AA$^2$ surface with adatoms randomly distributed on the surface. The concentration of adatoms on the surface is described by a coverage ($\theta$), given as the ratio of adatoms on the surface to the maximum acceptable number. The dynamic simulations were performed in the canonical ensemble, with a timestep of 0.5 fs, using the popular LAMMPS MD code.\cite{Plimpton_1}

We start by exploring the temporal evolution of a system with $\theta$ = 0.14 at 300 K. Snapshots during the course of the dynamic simulation are illustrated in Figure \ref{ripening_time}, up to a few nanoseconds. The randomly distributed adatoms quickly (in a few picoseconds) form small islands with a density $\approx$ 4-5 atoms, and once formed remain intact. This is consistent with past theoretical studies where dimers, trimers, and larger clusters were shown to be stable once formed and prefer to move in a concerted manner \cite{Stumpf_1, Bogicevic_2, Chang_1, Chang_2}. As island density increases, the mobility decreases, given the proportional number of bonds that need to be broken, significantly increasing the time between any relevant concerted displacements. Nevertheless, at 300 K the thermal energy is sufficient to overcome these barriers and the individual clusters ripen to form an island after 2 ns. Also, the island formed is primarily 2D, with no observable evaporation of adatoms (pathway (i) in Figure \ref{pathways}). 

\begin{figure}
	\centering
	\includegraphics[scale=0.27]{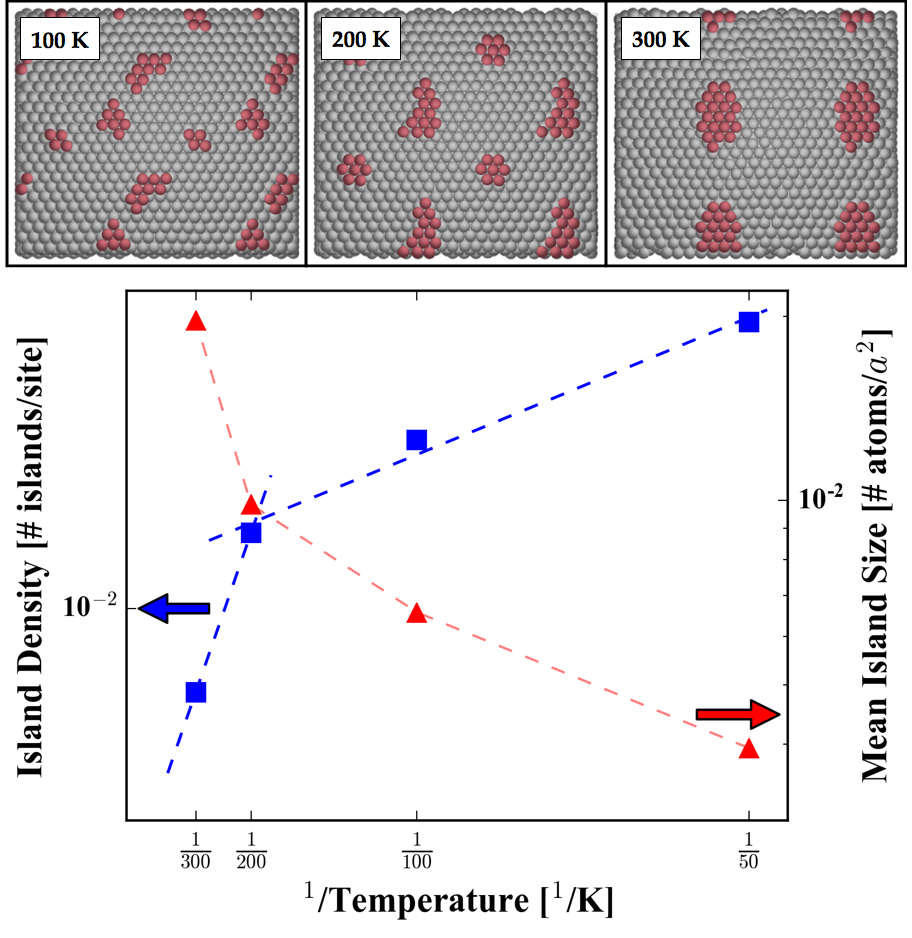}
	\caption{Top Panel: Island ripening as a function of temperature. Shown here is a simulation at the end of 2.5 ns for 100 K, 200 K, 300 K, at $\theta$ = 0.14. Grey and red colored atoms correspond to the surface atoms and adatoms, respectively. Bottom Panel: Island density and size as a function of temperature. Two scaling regimes are observed with a transition temperature of $\approx$ 200 K.} 
	\label{ripening_temperature}
\end{figure}

Temperature plays a critical role in the morphological shape and density of the islands formed. Past experimental studies, at $\theta$ = 0.11, by Busse \textit{et al.} reveals distinct structural patterns for the islands formed in a temperature range of 50 - 300 K.\cite{Busse_1,Busse_2} At low temperatures the lack of thermal energy results in hit and stick islands, that are formed by aggregation of nearby atoms to form several small clusters of islands, that transitions into compact islands as the temperature increases. To better understand the role of temperature, dynamical simulations between 50 K - 300 K were performed, as shown in Figure \ref{ripening_temperature} (top panel). At 100 K, we observe several small islands that remain immobile once formed. At 200 K, the islands grow in a dendritic fashion, whereby clusters of atoms group together with no significant rearrangement. The elementary processes, such as kink breaking, corner diffusion, etc., required to smoothen islands are activated only at high temperatures, and in the case of Al (111) beyond 250 K. These findings are consistent with past kinetic Monte Carlo results for other metal surfaces, whereby, fractal patterns dominate at low temperature, which transition into dendritic patterns, and culminate as compact islands as the temperature increases.\cite{Ratsch_1,Ovesson_1,Bogicevic_1} Note that the shape of islands observed in the MD simulations will be governed by the underlying reference theory used. It is known that a PBE level theory fails to correctly capture the anisotropy behavior of edge and corner diffusions, leading to triangular like rather than compact islands.\cite{Ovesson_1} Such deficiencies translate into our simulations as well. Nevertheless, the data-driven nature of AGNI force fields provides the flexibility of incorporating more accurate reference data to overcome such deficiencies. 

Futher, in the case of an Al (111) surface Busse \textit{et al.} experimentally observed two distinct growth regimes, as measured by the change in island density and temperature (T), one at T $<$ 200 K  and one above T $>$ 200 K. \cite{Busse_1}. We undertake a similar such analysis for the structures generated by the molecular dynamics simulations (c.f., bottom panel of Figure \ref{ripening_temperature}). An identical transition, consistent with the past experimental results, is observed at T $\approx$ 200 K. For the two regimes we compute the activation energy by measuring the slope, resulting in values of 1.2 K$^{-1}$ and 4.2 K$^{-1}$. The consistent features observed (stability of island shapes and growth regimes) suggests that the atomic forces predicted by AGNI force fields drive the dynamical evolution of the system in a manner consistent with both the thermodynamics and kinetics governing the process.


To summarize, in this work we demonstrated the use of machine learning force fields in studying the ripening phenomena of adatoms on an Al (111) surface, using MD simulations. Here, we confirm the two growth regimes observed experimentally in a temperature range of 50 - 300 K. At low temperatures the ripening phenomena is localized resulting in fractal like islands, and transitions into more compact islands as the temperature increases. This transition is observed at a temperature of $\approx$ 200 K, and is in excellent agreement with past experimental data. Clearly, the simulations undertaken by combining machine learning methods and quantum mechanical data demonstrates the fidelity in describing dynamical phenomena at larger length and time scales. The dependence on such hybrid methodologies will become increasingly more important, as we continually strive to push the envelope of atomistic modeling capabilities to more interesting materials phenomena, e.g., phase transformations or reactions on surfaces, all of which require going beyond a purely quantum mechanical description. 

This work was supported financially by the Office of Naval Research (Grant No. N00014-14-1-0098). The authors would like to acknowledge K. B. Lipkowitz for helpful discussions. Partial computational support through a Extreme Science and Engineering Discovery Environment (XSEDE) allocation is also acknowledged.

\bibliography{references}

\end{document}